\documentclass{ws-procs_arxiv}

\mathchardef\mhyphen="2D

\newcommand{\LCDM}{$\Lambda$CDM}

\newcommand{\org}{{\scshape origami}}

\newcommand{\citep}{\cite}
\newcommand{\citet}{\cite}

\chardef\til=`\~

\begin{document}

\title{ORIGAMI: DELINEATING COSMIC STRUCTURES WITH PHASE-SPACE FOLDS}

\author{MARK C.\ NEYRINCK$^{*\ddag}$, BRIDGET L.\ FALCK$^\dagger$ AND ALEX S.\ SZALAY$^*$}

\address{$^*$Dept of Physics and Astronomy, The Johns Hopkins University, Baltimore, MD 21218, USA\\
$^\dagger$Institute of Cosmology and Gravitation, University of Portsmouth, Portsmouth PO1 3FX, UK\\
$^\ddag$Email: neyrinck@pha.jhu.edu}

\begin{abstract}
Structures like galaxies and filaments of galaxies in the Universe come about from the origami-like folding of an initially flat three-dimensional manifold in 6D phase space.  The \org\ method identifies these structures in a cosmological simulation, delineating the structures according to their outer folds.  Structure identification is a crucial step in comparing cosmological simulations to observed maps of the Universe.  The \org\ definition is objective, dynamical and geometric: filament, wall and void particles are classified according to the number of orthogonal axes along which dark-matter streams have crossed.  Here, we briefly review these ideas, and speculate on how  \org\  might be useful to find cosmic voids.
\end{abstract}

\vspace{1mm}
\bodymatter

In general relativity, the 4D spacetime manifold is distorted by matter and energy.  Another manifold that pervades spacetime distorts and folds, as well: the dark-matter sheet.  This is the set of observers in free-fall for all cosmic time.  In GR, the dark-matter sheet is a spacelike surface in the synchronous gauge.

Often, dark-matter particles in a gravitational $N$-body simulation are thought of as physical blobs of mass.  But since simulation particles are dozens of orders of magnitude larger and more massive than physical dark-matter particles, it is more accurate to think of particles as vertices on a mesh\cite{ShandarinEtal2012,AbelEtal2012}, which distorts to represent density fluctuations.  Where there is a bit more matter than average, the mesh has contracted, and where there is less, the mesh has expanded.

The primordial state of the universe was well-characterized by particles separated from each other in position dimensions, but with vanishing bulk velocity.  In this sense, the matter sheet was flat.  A particle's Lagrangian, or initial, coordinates locate it within the sheet; its Eulerian coordinates are positions in the usual Newtonian-plus-cosmic-expansion approximation.

\begin{figure}
  \begin{minipage}[b]{0.55\linewidth}
    \centering
    \includegraphics[width=0.6\linewidth]{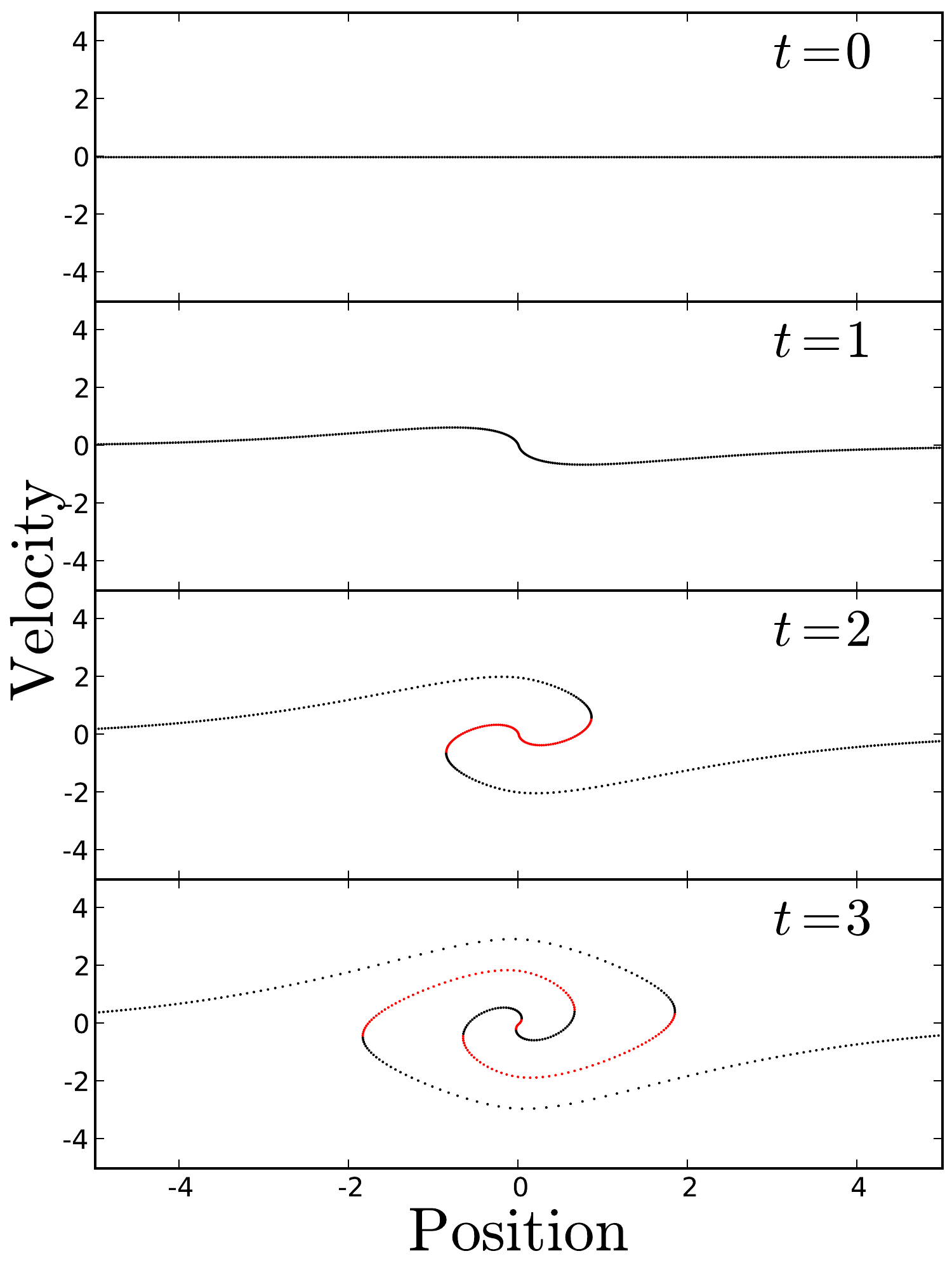}
  \end{minipage}
  \begin{minipage}[b]{0.35\linewidth}
        \centering

  \caption{A schematic phase-space spiral that corresponds to the
    collapse of a `halo' in a 1D universe.  Patches
    oriented forwards in position space (i.e. projecting down to the
    $x$-axis) are colored black, while patches oriented backwards are
    red.  These two possibilities make the set of streams (contiguous
    patches with the same orientation) two-colorable, even for 2D and 3D sheets.\cite{Neyrinck2012b}
    \label{fig:spirals}
    }
    \vspace{1cm}
  \end{minipage}
\end{figure}

In sufficiently overdense regions, the sheet bunches together and folds.
Dark-matter streams pass right through other streams, since dark matter is subject only to gravitational forces.  
Fig.\ \ref{fig:spirals} schematically shows collapse in a 1D universe
(with a 2D position-velocity phase space).  Initially, particles are almost equally
separated, but are drawn into the center, the Lagrangian string
winding up.  The two orientations of the initially flat string of
dark-matter particles are colored black (forward) and red (backward).
A stream is formally defined as a contiguous region on the string that is oriented the same way.  The boundaries between streams, folds in the
string when projected down to the $x$-axis, are called caustics.  Here, at the center of the 1D halo, multiple streams with different velocities occupy the same position.

\begin{figure}
  \begin{minipage}[b]{0.67\linewidth}
    \centering
    \includegraphics[width=\linewidth]{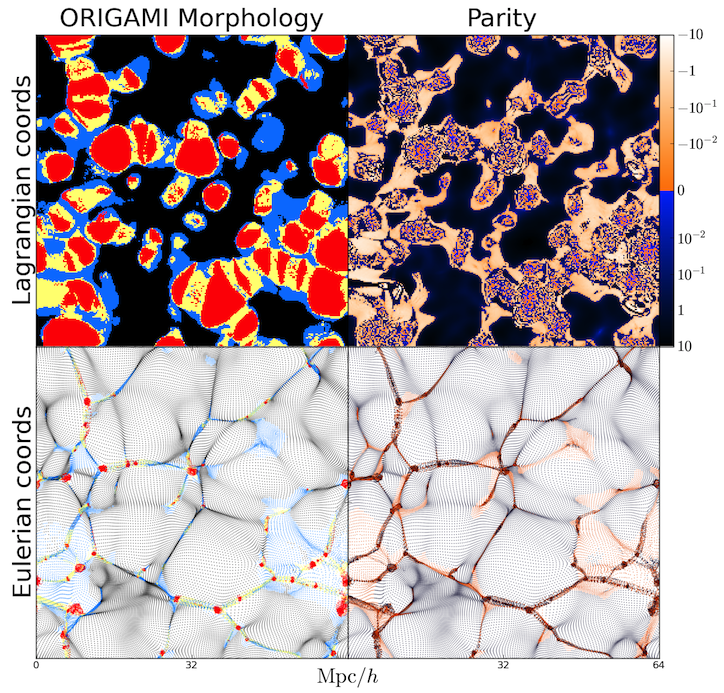}
  \end{minipage}
  \begin{minipage}[b]{0.3\linewidth}
        \centering
  \caption{The folding of a 2D sheet  from a 3D cosmological simulation.  Top, `unfolded' panels use Lagrangian coordinates; each pixel represents a particle.  Bottom, `folded-up' panels show particles in Eulerian $(x,y)$ coordinates.  At right, particles are colored
    according to their fluid elements' directed volumes (negative if a volume element has been inverted from caustic formation). 
    At left, void, wall, filament and halo
    \org\ morphologies are shown in black, blue, yellow and red,
    respectively.}
    \vspace{15mm}
      \label{fig:morphparity1}
  \end{minipage}
\end{figure}

Fig.\ \ref{fig:morphparity1} shows the folding up of a 2D
sheet of particles from a 3D cosmological simulation, with its
initial fluctuations smoothed compared to \LCDM\ for clarity. The top
panels show the sheet before folding; the boundaries between the light and dark regions at upper right show the creases along which folding occurs, detected by measuring the orientations/parities of fluid elements around particles.
Tracking creases gives roughly the outer boundaries of structures, but gives little information about what structures are built from these creases, e.g.\ their dimensionality/morphology.  

At left, structure dimensionalities have been detected for each particle with an algorithm called \org\ \cite{FalckEtal2012}.
In the one-dimensional halo of Fig.\ \ref{fig:spirals}, a natural boundary is where the number of streams changes from 1 to 3.  In three dimensions, structures are classified according to how
many perpendicular axes particles within them have been crossed along
by other particles.  Particles in voids, walls, filaments and haloes
have been crossed along 0, 1, 2, and 3 perpendicular directions.  This
is a parameter-free, objective, geometrical, and
dynamical identification of structures and placement of their
boundaries, that comes from considering the 6D phase space of both Lagrangian and Eulerian positions.  However, this simple particle-crossing criterion does not pick out substructures, which seem to require a more sophisticated, or insightful, algorithm to detect.

\begin{figure}
  \begin{minipage}[b]{0.5\linewidth}
    \centering
    \includegraphics[width=0.75\linewidth]{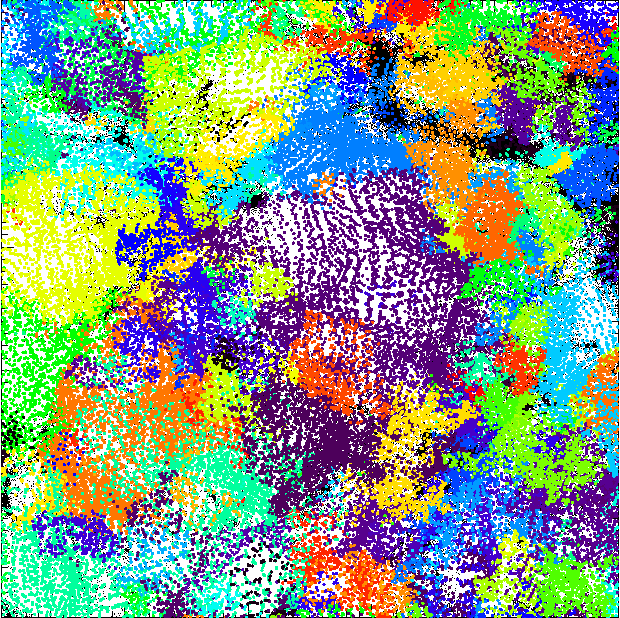}
  \end{minipage}
  \begin{minipage}[b]{0.45\linewidth}
        \centering
  \caption{A slice of an $N$-body simulation, showing a partition of no-stream-crossing particles into different(ly colored) voids.  The slice is $50\times 50\times 5$ $(h^{-1} {\rm Mpc})^3$.  Particles are linked to Voronoi neighbors until all particles belong to a void. We insist that each void have a single connected `core,' in which each Voronoi-tessellation particle density $<0.1$ times the mean.  Without such a criterion, almost all regions without stream-crossing in the simulation would link together.
    \label{fig:voidslice}
    }
  \end{minipage}
\end{figure}

In Ref.~\refcite{FalckEtal2012}, we showed the power of this approach to delineate haloes.  Here we give a preview of how the \org\ morphology tag delineates cosmic voids, i.e. whether voids can be usefully defined as regions surrounded by walls where caustics have formed.  Often, voids are thought of as nicely convex polyhedra, or even spheres, in cosmology.  But even at rather high resolution, we were surprised to see that many wall-like density ridges do not form complete caustics; there is essentially a single contiguous, percolating\cite{ShandarinEtal2004} no-stream-crossing region.  It seems that quite high resolution is necessary for caustics to carve out nice, convex voids.  Using a further density-based criterion, it is possible to partition the no-stream-crossing particles into roughly convex voids, as shown in Fig.\ \ref{fig:voidslice}.  We will report complete results in an upcoming paper\citet{FalckEtalPrep}.

\section*{Acknowledgment}
MCN is grateful for support from a New Frontiers in Astronomy and Cosmology grant from the Sir John Templeton Foundation.

\bibliographystyle{ws-procs975x65}
\bibliography{refs}

\end{document}